\documentclass[aps,prl,twocolumn,10pt,showpacs]{revtex4-1}
\usepackage{amsmath,amssymb,amsfonts}
\begin{document}
\title{Path Entropy Changes in Adiabatic Approximation}
\author{Jang-il Sohn}
\email{physicon@korea.ac.kr}
\affiliation{Department of Physics, Korea University, Seoul 136-713, Korea}
\date{September 26, 2012}
\begin{abstract}
By applying adiabatic theorem to a Markovian system, we calculate the adiabatic and diabatic entropy changes along a path.
As well known, the total path entropy change is separated into two parts, system and environment entropy changes, $\Delta S_{tot} = \Delta S_{sys} + \Delta S_{env}$.
The environment entropy change, $\Delta S_{env}$, is divided again into two parts, an adiabatic contribution due to work, $\Delta S_{\mathcal{W}}$, and a diabatic contributions due to heat, $\Delta S_{\mathcal{Q}}$.
In an adiabatic process, total path entropy change is same with the adiabatic path entropy change, $\Delta S_{A}$, which is given by sum of system entropy change and adiabatic contribution, $\Delta S_{A} = \Delta S_{sys} + \Delta S_{\mathcal{W}}$.
Mathematical form of $\Delta S_{A}$ is a type of excess heat entropy change, but $\Delta S_{A}$ is due to work.
By which, it is shown that the terms adiabatic and non-adiabatic contributions of $\Delta S_{na}$ and $\Delta S_{a}$ in [Phys. Rev. Lett. {\bf 104}, 090601 (2010)] should be completely switched, $i.e.$ $\Delta S_{na} \rightarrow \Delta S_{A}$ and $\Delta S_{a} \rightarrow \Delta S_{\mathcal{Q}}$ in fact.
\end{abstract}
\maketitle

{\it Introduction}.$-$
One of the attractions of fluctuation theorem (FT) is that they successfully describe the behavior of the second law of thermodynamics $\langle \Delta S \rangle \ge 0$ in microscopic limit.
Ever since the first FT is introduced by Evans, Cohen, and Morriss \cite{PhysRevLett.71.2401} in 1993, various FTs have been developed. Among them the FTs proposed by Esposito and Van den Broeck \cite{PhysRevLett.104.090601,PhysRevE.82.011143} are considered to be the most generalized ones.

In their study, Esposito and Van den Broeck have expressed the total entropy change for a single path as a sum of two distinctive contributions, $\Delta S_{na}$ and $\Delta S_{a}$, which are called respectively as \lq\lq{}non-adiabatic\rq\rq{} and \lq\lq{}adiabatic\rq\rq{} because of the (quantum) adiabatic theorem \cite{personal_email,griffiths1995introduction}.
Their generalization of FT is plausible, but their naming is misleading because both terms do not accurately reflect the entropy changes from non-adiabatic or adiabatic processes.

$\Delta S_{na}$ has an another name, the excess heat entropy change.
But, as seen in \cite{santillan2011irreversible}, the excess heat entropy change can be derived in adiabatic approximation.
Probably there is something wrong in \cite{PhysRevLett.104.090601}.
In order to conform which one is an adiabatic contribution, one must perform a rigorous verification in an adiabatic process (or in a non-adiabatic process), but they missed it.

In this work, we study adiabatic and dibatic entropy changes along a single path.
At first, the system is defined in {\it Model System}, and we will conjecture which one is the adiabatic path entropy change in {\it Thermodynamics Interpretations}.
Then the conjectures are verified in {\it Adiabatic Approximation} by dividing the underlying processes into adiabatic and diabatic processes.
In which, coarse graining and time scale separation \cite{{santillan2011irreversible},{PhysRevE.85.041125}} will be appropriate mathematical tools for that.
Lastly we conclude with discussions about excess heat entropy change and adiabatic approximation in {\it Conclusion}.

{\it Model System}.$-$
Imagine a Markovian system of $N$ possible states with an ergodic control parameter $\lambda_t$.
We assume that non-zero $\dot \lambda_t$ gives the system work added (or done), $\mathcal{W}$, and temperature of a reservoir (or reservoirs) is constant.
At any time $t$, the control parameter changes first from $\lambda_t$ to $\lambda_{t+\tau_0}$ for unit time $\tau_0$, and then the system goes from previous state $i_t$ to next state $i_{t+\tau_0}$ by time evolution operator $w_{i_{t+\tau_0}i_{t}}^{\tau_0}(\lambda_{t+\tau_0})$ which is controlled by schedule of a control parameter.
If the system undergoes a transition from a state to an another state, $w_{i_{t+\tau_0}i_{t}}^{\tau_0}(\lambda_{t+\tau_0})$ plays a role of a transition probability.
If the system is still in previous state, $i.e.$ if $i_{t} = i_{t+\tau_0}$, then $w_{i_{t+\tau_0}i_{t}}^{\tau_0}(\lambda_{t+\tau_0})$ plays a role of a waiting probability.

Let us consider a time evolution of the system which goes from an initial state $i_0$ to a final state $i_{T}$ along a path
\begin{equation}
\label{path}
[ \mathbf{\sigma} ]_{0}^{T} = [i,\lambda]_{0}^{T} \dot =
\left( i_0 \xrightarrow {\lambda_{\tau_0}} i_{\tau_0}
\xrightarrow {\lambda_{2\tau_0}} \cdots
\xrightarrow {\lambda_{T}} i_{T} \right)
\end{equation}
for a period of $T = n \tau_0$, where $\sigma$ is a pair of $i$ and $\lambda$.
At the end of the schedule, the system goes back along the reversed path.
For a single path, total entropy change is defined by
\begin{equation}
\label{total_entropy}
\Delta S_{tot}[\sigma]_{0}^{T} \equiv \ln
\frac{\mathcal{P}[\sigma]_{0}^{T}}{\mathcal{P}[\sigma]_{T}^{0}},
\end{equation}
where $f[\sigma]_{t_1}^{t_2}$ is a function of the path $[\sigma]_{t_1}^{t_2}$ \cite{PhysRevLett.104.090601,harris2007fluctuation}.
Here, $\mathcal{P}[\sigma]_{0}^{T}$ and $\mathcal{P}[\sigma]_{T}^{0}$ are respectively forward and reversed path probabilities, which are given by
\begin{equation}
\label{forward_path}
\mathcal{P}[\sigma]_{0}^{T} = \prod_{m=1}^{n}
w_{i_{m\tau_0}i_{(m-1)\tau_0}}^{\tau_0}(\lambda_{m\tau_0}) p_{i_0}(0),
\end{equation}
\begin{equation}
\label{reversed_path}
\mathcal{P}[\sigma]_{T}^{0} = \prod_{m=1}^{n}
w_{i_{(m-1)\tau_0}i_{m\tau_0}}^{\tau_0}(\lambda_{m\tau_0}) p_{i_T}(T)
\end{equation}
where $p_{i_0}(0)$ and $p_{i_T}(T)$ are respectively probabilities of the initial state $i_0$ and the final state $i_T$.
Since $w_{i_{t+\tau_0}i_{t}}^{\tau_0}(\lambda_{t+\tau_0})$ is not a transition probability but a time evolution operator, there is no need to consider waiting time in path probabilities.

{\it Thermodynamics Interpretations}.$-$
Here, we will study adiabatic and diabatic path entropy changes in the view of thermodynamics.

Total entropy change (\ref{total_entropy}) for a single path can be rewritten as
\begin{equation}
\label{total_path_entropy}
\Delta S_{tot}[\sigma]_{0}^{T} = \sum_{m=1}^{n} \ln \frac
{w_{i_{m\tau_0}i_{(m-1)\tau_0}}^{\tau_0}(\lambda_{m\tau_0}) p_{i_0}(0)}
{w_{i_{(m-1)\tau_0}i_{m\tau_0}}^{\tau_0}(\lambda_{m\tau_0}) p_{i_T}(T)}.
\end{equation}
As well known, it can be a sum of system and environment entropy changes, $\Delta S_{tot} = \Delta S_{sys} + \Delta S_{env}$ \cite{harris2007fluctuation}.
Those are respectively given by
\begin{equation}
\label{sys_ref}
\Delta S_{sys}[\sigma]_{0}^{T} = \ln  \frac{ p_{i_0}(0)} { p_{i_{T}}(T)}
\end{equation}
\begin{equation}
\label{env_ref}
\text{and }\text{ }
\Delta S_{env}[\sigma]_{0}^{T} = \sum_{m=1}^{n} \ln \frac
{w_{i_{m\tau_0}i_{(m-1)\tau_0}}^{\tau_0}(\lambda_{m\tau_0})}
{w_{i_{(m-1)\tau_0}i_{m\tau_0}}^{\tau_0}(\lambda_{m\tau_0})}.
\end{equation}

Reflecting on the first law of thermodynamics, $\Delta \mathcal{E} = \mathcal{W} + \mathcal{Q}$, the system can undergo a transition due to work $\mathcal{W}$ or heat $\mathcal{Q}$.
So, $\Delta S_{env}[\sigma]_{0}^{T}$ can be regarded as a sum of two contributions,
\begin{equation}
\label{two_contributions}
\Delta S_{env}[\sigma]_{0}^{T} = \Delta S_{\mathcal{W}}[\sigma]_{0}^{T} + \Delta S_{\mathcal{Q}}[\sigma]_{0}^{T},
\end{equation}
where $\Delta S_{\mathcal{W}}[\sigma]_{0}^{T}$ and $\Delta S_{\mathcal{Q}}[\sigma]_{0}^{T}$ are respectively an adiabatic contribution caused by $\mathcal{W}$ and a diabatic contribution caused by $\mathcal{Q}$ (which are quitely different from Esposito and Van den Broeck\rq{}s adiabatic and non-adiabatic contributions).

According to the adiabatic theorem, the system becomes an adiabatic process when $\lambda_t$ varies very slowly \cite{{griffiths1995introduction},{santillan2011irreversible},{PhysRevE.85.041125}}.
But there is one more requirement, time scale separation, $\tau_s$.
If the unit time or coarse graining period are larger than $\tau_s$, the system goes to an adiabatic process \cite{santillan2011irreversible}.
If not, even $\lambda_t$ varies very slowly, the system is a diabatic process, thereby there is a non-zero entropy change due to $\mathcal{Q}$ in non-equilibrium steady state.
The non-zero entropy change is, as well known, the house-keeping heat entropy change \cite{oono1998steady,speck2005integral,harris2007fluctuation} which is given by
\begin{equation}
\label{Q_entropy}
\Delta S_{\mathcal{Q}}[\sigma]_{0}^{T} =
\sum_{m=1}^{n} \ln \frac
{w_{i_{m\tau_0}i_{(m-1)\tau_0}}^{\tau_0}(\lambda_{m\tau_0})p_{i_{(m-1)\tau_0}}^{st}(\lambda_{m\tau_0})}
{w_{i_{(m-1)\tau_0}i_{m\tau_0}}^{\tau_0}(\lambda_{m\tau_0})p_{i_{m\tau_0}}^{st}(\lambda_{m\tau_0})},
\end{equation}
where $p_{i}^{st}(\lambda)$ is steady state probability for $\lambda$.
In \cite{PhysRevLett.104.090601}, they have called it as an adiabatic contribution, $\Delta S_{a}$, but which is a contradiction in the view of thermodynamics, because $\Delta S_{\mathcal{Q}}$ (or $\Delta S_{a}$) is definitely due to $\mathcal{Q}$.
Additionally, if the system is coarse grained for a period $\tau$ larger than $\tau_s$, the system indeed goes to an adiabatic process as mentioned above, thereby $\Delta S_{\mathcal{Q}}$ vanish \cite{santillan2011irreversible}.
Therefore, $\Delta S_{\mathcal{Q}}$ (or $\Delta S_{a}$) in non-equilibrium steady state can not be an adiabatic contribution at all, but rather a diabatic contribution.

On the other hand, $\Delta S_{\mathcal{W}}$ is definitely an adiabatic contribution, since non-zero $\dot \lambda_t$ gives the system $\mathcal{W}$ as defined in {\it Model System}.
From (\ref{env_ref}), (\ref{two_contributions}) and (\ref{Q_entropy}), $\Delta S_{\mathcal{W}}$ is written as
\begin{equation}
\label{adiabatic_contribution}
\Delta S_{\mathcal{W}}[\sigma]_{0}^{T} =
\sum_{m=1}^{n} \ln \frac{p_{i_{m\tau_0}}^{st}(\lambda_{m})}
{p_{i_{(m-1)\tau_0}}^{st}(\lambda_{m})}.
\end{equation}
Here, $\Delta S_{\mathcal{W}}$ is a type of excess heat entropy change \cite{santillan2011irreversible,PhysRevLett.86.3463}, but different from excess heat entropy change (that will be discussed in {\it Conclusion}).

In summary up to here, total entropy change along a single path can be a sum of three different contributions,
\begin{equation}
\Delta S_{tot}[\sigma]_{0}^{T} = \Delta S_{sys}[\sigma]_{0}^{T} + \Delta S_{\mathcal{W}}[\sigma]_{0}^{T} + \Delta S_{\mathcal{Q}}[\sigma]_{0}^{T}.
\end{equation}
If there is no house-keeping heat entropy change, the system is an adiabatic process, thereby path entropy change should be $\Delta S_{A} = \Delta S_{sys} + \Delta S_{\mathcal{W}}$.
However, in a diabatic process, $\Delta S_{\mathcal{Q}}$ is also to be considered.
Therefore total path entropy change should be divided into an adiabatic path entropy change and a diabatic contribution, which are respectively
\begin{equation}
\label{adiabatic_thermo}
\Delta S_{A}[\sigma]_{0}^{T} =
\ln  \frac{ p_{i_0}(0)} { p_{i_{T}}(T)} +
\sum_{m=1}^{n} \ln \frac{p_{i_{m\tau_0}}^{st}(\lambda_{m})}
{p_{i_{(m-1)\tau_0}}^{st}(\lambda_{m})}
\end{equation}
\begin{equation}
\label{diabatic_thermo}
\Delta S_{\mathcal{Q}}[\sigma]_{0}^{T} =
\sum_{m=1}^{n} \ln \frac
{w_{i_{m\tau_0}i_{(m-1)\tau_0}}^{\tau_0}(\lambda_{m\tau_0})p_{i_{(m-1)\tau_0}}^{st}(\lambda_{m})}
{w_{i_{(m-1)\tau_0}i_{m\tau_0}}^{\tau_0}(\lambda_{m\tau_0})p_{i_{m\tau_0}}^{st}(\lambda_{m})}.
\end{equation}
By comparing (\ref{adiabatic_thermo}) and (\ref{diabatic_thermo}) with \cite{PhysRevLett.104.090601}, it is clear that $\Delta S_{na} = \Delta S_{A}$ and $\Delta S_{na} = \Delta S_{\mathcal{Q}}$.
Therefore, $\Delta S_{na}$ must be corrected to an {\it adiabatic entropy change}, and $\Delta S_{a}$ is related to a {\it diabatic contribution}, if our conjectures, (\ref{adiabatic_thermo}) and (\ref{diabatic_thermo}), are right.

{\it Adiabatic Approximation}.$-$
Now, our conjectures above is verified by calculating path entropy changes in adiabatic approximation (or quasi-static approximation).

According to the adiabatic theorem, the system becomes an adiabatic process when an external field varies very slowly \cite{{griffiths1995introduction},{santillan2011irreversible},{PhysRevE.85.041125}}.
It is well known that the adiabatic process corresponds to a quasi-static process in thermodynamics.
Although the quasi-static process means a process to make the system staying extremely close to equilibrium steady states, it could be also applied to a non-equilibrium thermodynamics system \cite{oono1998steady}.
So, performing calculations in the quasi-static process will give an adiabatic entropy change.

At here, we have to deliberate on the adiabatic theorem in thermodynamics.
As mentioned above in {\it Thermodynamics Interpretations}, 
in order to make the system an adiabatic process,
the follows are to be required:
Firstly $\lambda_t$ should be slowly changing, and secondly unit time or coarse graining period have to be larger than the time scale separation.
In fact, the second requirement contains the first one.
So, if a process is coarse grained for a period $\tau = n_0 \tau_0$ larger than $\tau_s$, the system becomes an adiabatic process \cite{{santillan2011irreversible},{PhysRevE.85.041125}} (we do not study in detail about $\tau_s$ in this work).
Hence, when $\tau \gg \tau_0$, we can calculate an adiabatic entropy change exactly.
The diabatic entropy change can be obtained as an additional part when $\tau \simeq \tau_0$ by comparing the total entropy change with the adiabatic entropy change.

Let us introduce our strategy to calculate $\Delta S_{\mathcal{W}}$ in a Markovian system.
Firstly, introducing a temporarily time homogeneous Markov process, the time evolution operator is coarse grained for a period $\tau = n_0 \tau_0$.
Secondly, time evolution operators corresponding to respectively adiabatic and diabatic transitions, $\mathbf{W}^{st}(\lambda_t)$ and $\mathbf{N}^{\tau}(\lambda_t)$, are defined.
Lastly, an adiabatic entropy change, $\Delta S_{A}$, is calculated in adiabatic limit (or quasi-static limit).

The first step starts from here.
In the system, Markov chain at time $t$ for unit time $\tau_0$ is given by
\begin{equation}
\label{Markov_chain}
\mathbf{P}(t+\tau_0) = \mathbf{W}^{\tau_0}(\lambda_{t+\tau_0}) \mathbf{P}(t),
\end{equation}
where $\mathbf{P}(t) = (p_1(t), \cdots, p_N(t))^{Transpose}$ is a column vector of a state distribution, and $\mathbf{W}^{\tau_0}(\lambda_{t}) = \left( w_{ij}^{\tau_0}(\lambda_{t}) \right)$ is an $N \times N $ matrix of a time evolution operator.
At this step, we consider a partial path
\begin{equation}
\label{path_diagram_explicit}
[\sigma]_{t}^{t+\tau} \dot =
\left( 
i_t \xrightarrow {\lambda_{t+\tau_0}} i_{t+\tau_0}
\xrightarrow {\lambda_{t+2\tau_0}} \cdots
\xrightarrow {\lambda_{t+\tau}} i_{t+\tau} 
\right),
\end{equation}
for coarse graining period $\tau = n_0 \tau_0$ along which (\ref{Markov_chain}) is rewritten as
\begin{equation}
\label{Markov_chain_complicated}
\mathbf{P}(t+\tau) =
\underbrace{\mathbf{W}^{\tau_0}(\lambda_{t+\tau}) \cdots
\mathbf{W}^{\tau_0}(\lambda_{t+\tau_0})}_{n_0=\tau/\tau_0} \mathbf{P}(t).
\end{equation}
But it is rather complicated to calculate the matrices, because $\lambda_t$ varies in time.

To make it simple, the temporarily time homogeneous Markov process is introduced.
When the control parameter is fixed at $\lambda_{t+\tau}$, the system will evolve along the following partial path
\begin{equation}
\label{partial_path}
[\sigma]_{t}^{t+\tau} \dot =
\left( i_t \xrightarrow {\lambda_{t+\tau}} i_{t+\tau_0}
\xrightarrow {\lambda_{t+\tau}} \cdots
\xrightarrow {\lambda_{t+\tau}} i_{t+\tau} \right).
\end{equation}
In this partial path,
(\ref{Markov_chain_complicated}) is simplified as
\begin{equation}
\label{Markov_chain_in_tau}
\mathbf{P}(t+\tau) = \mathbf{W}^\tau(\lambda_{t+\tau}) \mathbf{P}(t),
\end{equation}
where
\begin{equation}
\label{Markov_matrix_in_tau}
\mathbf{W}^{\tau}(\lambda_{t+\tau}) \equiv
\underbrace{\mathbf{W}^{\tau_0}(\lambda_{t+\tau})
\cdots \mathbf{W}^{\tau_0}(\lambda_{t+\tau})}_{n_0=\tau/\tau_0}
= \left( w_{ij}^{\tau}(\lambda_{t+\tau})  \right).
\end{equation}
The coarse grained time evolution operator $w^{\tau}_{ij}(\lambda_{t+\tau})$ looks like a singular jump from a previous state to a next state,
but obviously contains all transition and all waiting because it is a result of a path integration for all possible partial path,
$w^{\tau}_{ij}(\lambda_{t+\tau})
= \sum_{[\lambda]} w_{ij}^{\tau}[\lambda]_{t}^{t+\tau}$
where $w_{ij}^{\tau}[\lambda]_{t}^{t+\tau} = w_{ii_{t+\tau-\tau_0}}^{\tau_0}(\lambda_t) \cdots w_{i_{t+\tau_0}j}^{\tau_0}(\lambda_t) $ in which transitions (or waitings) in the middle of the partial path (\ref{partial_path}) are just veiled.

Here is the second step.
If $\tau \rightarrow \infty$ or $\lambda_t = \lambda = const$, then the system heads to a unique steady sate and finally reaches to there regardless of an initial state $\mathbf{P}(0)$,
\begin{equation}
\label{quasi_static_Markov_chain}
\mathbf{P}^{st}(\lambda) = \mathbf{W}^{st}(\lambda) \mathbf{P}(0).
\end{equation}
where the steady state time evolution operator is
\begin{equation}
\label{Wst_def}
\mathbf{W}^{st}(\lambda) \equiv \mathbf{W}^{\infty}(\lambda).
\end{equation}
Here, $\mathbf{P}^{st}(\lambda) = \left( p^{st}_{1}(\lambda),
\cdots, p^{st}_{N}(\lambda) \right)^{Transpose}$
is the unique steady state distribution for $\lambda$.
Since the equation (\ref{quasi_static_Markov_chain}) holds always, a trial distribution, $\mathbf{P}' = \left( \cdots,  0, p'_{k} = 1 , 0 , \cdots  \right)^{T}$ for any $k$, might be chosen in the place of $\mathbf{P}(0)$.
Then, from (\ref{quasi_static_Markov_chain}) and $\mathbf{P}'$, the following relation is derived,
\begin{equation}
\label{st_qs_relation}
p^{st}_{i}(\lambda) = w^{st}_{ik}(\lambda)
\end{equation}
for all $i$ and $k$.
For an any ergodic system, the relation is indeed valid and already known in
mathematical fields of stochastic processes \cite{ross1996stochastic}.
The relation means that all column vectors of $\mathbf{W}^{st}(\lambda)$
are same with $\mathbf{P}^{st}(\lambda)$,
\begin{equation}
\label{quasi_static_and_principal_eigen_vector}
{\mathbf W}^{st}(\lambda)
= \left( \begin{array}{cccc}
p^{st}_{1}(\lambda) & p^{st}_{1}(\lambda) & \cdots & p^{st}_{1}(\lambda) \\
p^{st}_{2}(\lambda_{t}) & p^{st}_{2}(\lambda) & \cdots & p^{st}_{2}(\lambda) \\
\vdots & \vdots & \ddots & \vdots \\
p^{st}_{N}(\lambda) & p^{st}_{N}(\lambda) & \cdots & p^{st}_{N}(\lambda)
\end{array} \right).
\end{equation}
To obtain this matrix, there is no need to multiply $\mathbf{W}^{\tau_0}(\lambda)$ infinitely.
That is simply obtained by the principal eigen-value equation,
$\mathbf{P}^{st}(\lambda) =
\mathbf{W}^{\tau}(\lambda) \mathbf{P}^{st}(\lambda)$.

Now, let us consider a whole path from initial state $i_0$ to final state $i_T$,
\begin{equation}
\label{path_diagram_0_T}
[\sigma]_{0}^{T} \dot =
\left( i_0 \xrightarrow {\lambda_{0}} i_{\tau}
\xrightarrow {\lambda_{\tau}} \cdots
\xrightarrow {\lambda_{T-\tau}} i_{T} \right),
\end{equation}
which is a sequence of the coarse grained partial paths (\ref{partial_path}), where $T=n\rq{}\tau$.
When $\tau \gg \tau_0$ or $\lambda_t$ varies very slowly comparing to $\tau$, the system goes to adiabatic limit, and the time evolution operator becomes a steady state time evolution operator,
\begin{equation}
\label{matrix_in_adiabatic_limit}
\mathbf{W}^{\tau}(\lambda_t) \simeq \mathbf{W}^{st}(\lambda_t).
\end{equation}
In this limit, the system is evolving near steady states of $\lambda_t$,
$i.e$ quasi-static processes (in thermodynamics) or adiabatic process (in quantum mechanics).
On the other hand, if $\tau \simeq \tau_0$ or $\lambda_t$ varies rapidly comparing to $\tau$, (\ref{matrix_in_adiabatic_limit}) does not hold anymore, and we have to consider an additional part corresponding to a diabatic contribution which is given by definition, $\mathbf{N}^{\tau}(\lambda_t)  \equiv \mathbf{W}^{\tau}(\lambda_t) - \mathbf{W}^{st}(\lambda_t)$.
So, the time evolution matrix can be separated into two parts,
\begin{equation}
\label{separated_transition_matrix}
\mathbf{W}^{\tau}(\lambda_t) =
\mathbf{W}^{st}(\lambda_t) + \mathbf{N}^{\tau}(\lambda_t).
\end{equation}

This is the last step to calculate the path entropy changes.
In adiabatic limit, $i.e.$ when  $\tau \gg \tau_0$ or $\dot \lambda_t$ is very small comparing to $\tau$, total path entropy change is same with adiabatic path entropy change. 
So, from (\ref{total_path_entropy}) and (\ref{st_qs_relation}) and (\ref{matrix_in_adiabatic_limit}), it is simply obtained as
\begin{equation}
\label{adiabatic_adia}
\Delta S_{ad}[\sigma]_{0}^{T} =
\ln \frac{p_{i_0}(0)}{ p_{i_T}(T)} +
\sum_{m=1}^{n'} \ln \frac
{p^{st}_{i_{m\tau}}(\lambda_{m\tau}) }
{p^{st}_{i_{(m-1)\tau}}(\lambda_{m\tau})}.
\end{equation}
On the other hand, in diabatic limit, $i.e.$ if $\tau \simeq \tau_0$ or $\lambda_t$ varies very rapidly comparing to $\tau$, we have to consider a diabatic contribution in entropy change which is given by definition, $\Delta S_{di} \equiv \Delta S_{tot} - \Delta S_{ad}$.
From (\ref{total_path_entropy}) and (\ref{adiabatic_adia}), the diabatic contribution is written as
\begin{equation}
\label{diabatic_adia}
\Delta S_{di}[\sigma]_{0}^{T} =
\sum_{m=1}^{n'} \ln \frac
{w_{i_{m\tau}i_{(m-1)\tau}}^{\tau}(\lambda_{m\tau})p_{i_{(m-1)\tau}}^{st}(\lambda_{m})}
{w_{i_{(m-1)\tau}i_{m\tau}}^{\tau}(\lambda_{m\tau})p_{i_{m\tau}}^{st}(\lambda_{m})}.
\end{equation}
By comparing (\ref{adiabatic_adia}), (\ref{diabatic_adia}) with (\ref{adiabatic_thermo}), (\ref{diabatic_thermo}), we can ascertain that
\begin{equation}
\Delta S_{ad}[\sigma]_{0}^{T} = \Delta S_{\mathcal{A}}[\sigma]_{0}^{T}
\text{ and }
\Delta S_{di}[\sigma]_{0}^{T} = \Delta S_{\mathcal{Q}}[\sigma]_{0}^{T}
\end{equation}
when $\tau = \tau_0$.
Therefore, our conjectures in {\it Thermodynamics Interpretation}, (\ref{adiabatic_thermo}) and (\ref{diabatic_thermo}), are right.

{\it Conclusion}.$-$
In this study, we conjectured which one is an adiabatic entropy change in {\it Thermodynamics Interpretation}, and then in {\it Adiabatic Approximation} we verified our conjectures: $\Delta S_{A}$ is an adiabatic entropy change, and $\Delta S_{\mathcal{Q}}$ is a diabatic contribution.
Therefore the terms of $\Delta S_{na}$ and $\Delta S_{a}$ in \cite{PhysRevLett.104.090601} should be completely switched as $\Delta S_{na} \rightarrow \Delta S_{A}$ and $\Delta S_{a} \rightarrow \Delta S_{\mathcal{Q}}$ in fact.

Even though the mathematical form of $\Delta S_{A}$ is a type $\Delta S_{\mathcal{Q}_{ex}}$ \cite{oono1998steady,PhysRevLett.86.3463}, $\Delta S_{A}$ is definitely different from $\Delta S_{\mathcal{Q}_{ex}}$ because their origins are different from each other.
In this work, because non-zero $\dot \lambda_t$ gives the system only $\mathcal{W}$, $\Delta S_{\mathcal{W}}$ is not excess heat entropy change definitely but a type of excess heat entropy change.
If non-zero $\dot \lambda_t$ gives the system $\mathcal{Q}$, $\Delta S_{\mathcal{Q}_{ex}}$ can be obtained as a diabatic entropy contribution, $i.e.$ $\Delta S_{\mathcal{Q}} = \Delta S_{\mathcal{Q}_{hk}} + \Delta S_{\mathcal{Q}_{ex}}$.

This work is consistent with other previous findings.
One of them is M. Santill{\'a}n and H. Qian's finding \cite{santillan2011irreversible}.
They obtained a similar result, $\dot S = - \mathcal{Q}_{ex} / T$, in adiabatic approximation.
In their study, they obtained the result in a molecular system with constant temperature, so that is a entropy change due to non-zero $\dot \lambda_t$ and a type of excess heat entropy change.
Moreover, T. Hatano's research \cite{PhysRevE.60.R5017} in 1999 can be an evidence supporting our results, too.
He verified the Jarzynski equality \cite{PhysRevLett.78.2690} without consideration of house-keeping heat \cite{PhysRevE.60.R5017}.
Remembering that the Jarzynski equality (or the Crooks relation \cite{crooks1998nonequilibrium}) holds originally in adiabatic limit and not related with $\mathcal{Q}$, T. Hatano\rq{}s finding in \cite{PhysRevE.60.R5017} can be regarded as performed in adiabatic limit, and indeed performed in adiabatic limit (or quasi-static limit).
Therefore, we conclude that $\Delta S_{A} = \Delta S_{sys} + \Delta S_{\mathcal{W}}$ is indeed an adiabatic entropy change, and $\Delta S_{\mathcal{Q}} = \Delta S_{\mathcal{Q}_{ex}} + \Delta S_{\mathcal{Q}_{hk}}$ is a diabatic contribution.

Lastly, we faced with a interesting topic which is the time scale separation between equilibrium and non-equilibirum domains.
Looking at (\ref{matrix_in_adiabatic_limit}), we can ascertain that detailed balance condition is satisfied in adiabatic limit or in slow process which is already reported in \cite{PhysRevE.83.041129}.
Researching about what determines $\tau_s$ can be an interesting topic.

{\it Acknowledgment}.$-$
I would like to thank Prof.s M.-S. Choi of Korea University, H. Hinrichsen of University of W\"{u}rzburg, C. Kwon, H. Park, Ph.D. S.K. Baek and statical physics group meeting members of Korea Institute of Advanced Study for discussions and comments, and especially Prof.s K. Goh, I.-M. Kim of Korea University for advices and supports.
This works is supported by Basic Science Research Program through NRF grant
funded by MEST (No. 2011-0014191).
\bibliographystyle{apsrev4-1}

\end{document}